\documentclass[aps,pra,showpacs,superscriptaddress,preprint]{revtex4}
\usepackage{graphicx}

\catcode`ð=\active
 \defð{\u{g}}
 \catcode`Ð=\active
 \defÐ{\u{G}}
 \catcode`Ý=\active
 \defÝ{\. I}
 \catcode`ö=\active
 \defö{\"{o}}
 \catcode`Ö=\active
 \defÖ{\"O}
 \catcode`ü=\active
 \defü{\"{u}}
 \catcode`Ü=\active
 \defÜ{\"{U}}
 \catcode`Þ=\active
 \defÞ{\c{S}}
 \catcode`þ=\active
 \defþ{\c{s}}
 \catcode`ý=\active
 \defý{{\i}}
 \catcode`ç=\active
 \defç{\d{c}}
 \catcode`Ç=\active
 \defÇ{\d{C}}

\begin{document}

\title{Solutions of Dirac Equation for Symmetric Generalized Woods-Saxon
Potential by the Hypergeometric Method }

\author{\small Sameer M. Ikhdair}
\email[E-mail: ]{sikhdair@neu.edu.tr}\affiliation{Department of
Physics, Near East University, Nicosia, Cyprus, Mersin-10, Turkey}
\author{\small Ramazan Sever}
\email[E-mail: ]{sever@metu.edu.tr}\affiliation{Department of
Physics, Middle East Technical  University, 06800, Ankara,Turkey}

\date{\today}

\begin{abstract}
The Dirac equation is solved approximately for the Hulthen
potential with the pseudospin symmetry for any spin-orbit quantum
number $\kappa$ in the position-dependent mass background.
Solutions are obtained reducing the Dirac equation into a
Schr\"{o}dinger-like differential equation by using an appropriate
coordinate transformation. The Nikiforov-Uvarov method is used in
the calculations to get energy eigenvalues and the corresponding
wave functions.
\end{abstract}

\maketitle

\newpage

\section{Introduction}

In the past few years there has been considerable work on
non-Hermitian Hamiltonians. Among this kind of Hamiltonians, much
attention
has been focused on the investigation of properties of so-called ${\cal PT}$%
-symmetric Hamiltonians. Following the early studies of Bender
{\it et al}. [1], the ${\cal PT}{\rm -}$symmetry formulation has
been successfuly utilized by many authors [2-8]. The ${\cal
PT}$-symmetric but non-Hermitian potentials could have real
spectra even it is non-Hermitian. Non-Hermitian Hamiltonians with
real or complex spectra have also been analyzed by using different
methods [3-6,9]. Non-Hermitian but ${\cal PT}$-symmetric models
have applications in different fields, such as nuclear physics
[10], condensed matter [11] and population biology [12].

When a particle is in a strong potential field, the relativistic
effect must be considered, which gives the correction for
non-relativistic quantum mechanics [13]. Taking the relativistic
effects into account, a particle in a potential field should be
described with the Klein-Gordon (KG) and Dirac equations. In
recent years, there has been an increased interest in finding
exact solutions to Schr\"{o}dinger, K-G, Dirac and Salpeter
equations for various potential schemes [13--27]. The problems
that can be exactly solved for the KG and/or Dirac equations are
seldom except a few examples, such as hydrogen atom and electrons
in a uniform magnetic field. Recently some
authors solved such relativistic equations for some potentials. \c{S}im\c{s}%
ek and E\u{g}rifes [15] have presented the bound-state solutions
of the one-dimensional ($1D$) Klein-Gordon (KG) equation for
${\cal PT}$-symmetric potentials with real and complex generalized
Hulth\'{e}n potential. Moreover, E\u{g}rifes and Sever [16]
investigated the bound state solutions of the $1D$ Dirac equation
with ${\cal PT}$-symmetric real and complex forms of generalized
Hulth\'{e}n potential. Yi {\it et al}. [19] obtained the energy
equations in the KG theory with equally mixed vector and scalar
Rosen-Morse-type potentials. In recent works, we have solved the spinless $%
1D $ Salpeter equation analytically for its exact bound state
spectra and wavefunctions with real and complex forms of the
${\cal PT}$-symmetric generalized Hulth\'{e}n potential [20]. We
have also investigated the bound state solutions of the $1D$ KG
equation with real and complex forms of the generalized
Woods-Saxon (WS) potential [21].

In the present work, we present a new procedure to construct
solution to the (1+1) dimensional Dirac equation with gauge
invariant (minimal) vector coupling. Using the Nikiforov-Uvarov
(NU) method [28], we are set to obtain the bound states solutions
(relativistic energy spectrum and the two-component spinor wave
functions) for particles trapped in a spherically
symmetric, generalized Woods-Saxon potential [21,22] which possesses a $%
{\cal PT}$-symmetry as well. Changing the values of the potential
parameters
from real to pure imaginary, we obtain Hamiltonians that may or may not be $%
{\cal PT}$-symmetric. Further, by making the same parameter change
in the energy spectrum and two-components spinor wave functions,
we obtain solutions for the new trigonometric and periodic
potential (real or complex) forms. The corresponding energy
spectra are either real or complex. The Dirac problem with a
${\cal PT}$-symmetric or a non-${\cal PT}$-symmetric imaginary
generalized Woods-Saxon (WS) potential is mapped into the exactly
solvable problem in which one may apply the ${\rm NU}$ method to
generate possible real energy spectra [28].

The paper is structured as follows: In Section \ref{THM}, we
briefly introduce the basic concepts of the hypergeometric method
(${\rm NU}$). Section \ref{S} is devoted to the solution of the
Dirac problem to obtain the exact bound state energy spectrum for
real and complex cases of generalized WS potentials and the lower
and upper spinor components eigenfunction by applying the NU
method. The results of this relativistic study are discussed in
Section \ref{RAC}.

\section{The Hypergeometric Method}

\label{THM}The basic equations of the theoretical background are
similar to those given in [15,16,20-25]. According to a breif
description of this method [28], the Dirac equation can be
transformed to the following generalized equation of
hypergeometric type after employing an appropriate coordinate
transformation, $s=s(r),$
\begin{equation}
\psi _{n}^{\prime \prime }(s)+\frac{\widetilde{\tau }(s)}{\sigma
(s)}\psi _{n}^{\prime }(s)+\frac{\widetilde{\sigma }(s)}{\sigma
^{2}(s)}\psi _{n}(s)=0,
\end{equation}%
where $\sigma (s)$ and $\widetilde{\sigma }(s)$ are polynomials,
at most of second-degree, and $\widetilde{\tau }(s)$ is a
first-degree polynomial. In order to find a particular solution to
Eq. (1), we use the following wave function

\begin{equation}
\psi _{n}(s)=\phi _{n}(s)y_{n}(s).
\end{equation}
This reduces Eq. (1) to an equation of a hypergeometric type

\begin{equation}
\sigma (s)y_{n}^{\prime \prime }(s)+\tau (s)y_{n}^{\prime
}(s)+\lambda y_{n}(s)=0,
\end{equation}
which demands that the following conditions be satisfied:

\begin{equation}
\sigma (s)=\pi (s)\frac{\phi (s)}{\phi ^{\prime }(s)},
\end{equation}

\begin{equation}
\tau (s)=\widetilde{\tau }(s)+2\pi (s),\text{ }\tau ^{\prime
}(s)<0,
\end{equation}
and $\lambda $ is thus a new eigenvalue equation for the
second-order differential equation becomes
\begin{equation}
\lambda =\lambda _{n}=-n\tau ^{\prime }(s)-\frac{n\left( n-1\right) }{2}%
\sigma ^{\prime \prime }(s),\text{ \ \ \ \ \ \ }n=0,1,2,....
\end{equation}
The polynomial $\tau (s)$ with the parameter $s$ and prime factors
show the
differentials at first degree be negative. It is worthwhile to note that $%
\lambda $ or $\lambda _{n}$ are obtained from a particular
solution of the
form $y(s)=y_{n}(s)$ which is a polynomial of degree $n.$ The other part $%
y_{n}(s)$ of the wavefunction (2) is the hypergeometric-type
function whose polynomial solutions are given by Rodrigues
relation

\begin{equation}
y_{n}(s)=\frac{B_{n}}{\rho (s)}\frac{d^{n}}{ds^{n}}\left[ \sigma
^{n}(s)\rho (s)\right] ,
\end{equation}
where $B_{n}$ is the normalization constant with the weight
function $\rho (s)$ must satisfy the following condition

\begin{equation}
w^{\prime }(s)-\left( \frac{\tau (s)}{\sigma (s)}\right) w(s)=0,\text{ }%
w(s)=\sigma (s)\rho (s).
\end{equation}%
On the other hand, in order to find the eigenfunctions, $\phi _{n}(s)$ and $%
y_{n}(s)$ in Eqs. (4) and (7) and eigenvalues $\lambda _{n}$ in
Eq. (6)$,$ we need to calculate the functions:
\begin{equation}
\pi (s)=\frac{\sigma ^{\prime }(s)-\widetilde{\tau }(s)}{2}\pm
\sqrt{\left(
\frac{\sigma ^{\prime }(s)-\widetilde{\tau }(s)}{2}\right) ^{2}-\widetilde{%
\sigma }(s)+k\sigma (s)},
\end{equation}%
\begin{equation}
\lambda =k+\pi ^{\prime }(s).
\end{equation}%
In principle, since $\pi (s)$ has to be a polynomial of degree at
most one, the expression under the square root sign in Eq. (9) can
be arranged to be the square of a polynomial of first degree [28].
This is possible only if its discriminant is zero. Thus, the
equation for $k$ obtained from the solution of Eq. (9) can be
further substituted in Eq. (10). In addition, the
energy eigenvalues are obtained from Eqs. (6) and (10).

\section{Solutions of the Quantum System}

\label{S}The exact solution of the relativistic wave equations is
of much concern in quantum mechanics. Many works appeared in the
recent years in that direction towards obtaining exact solution of
some relativistic wave equations for certain potentials of
physical interest (cf. [15,16,20-22] and references therein). For
a spinless particle of rest mass $m$ and total energy $E_{nl},$
the $1D$ time-independent Dirac equation with any given
interaction potential $V(x)$ in the vector coupling scheme
(choosing the natural atomic units $\hbar =c=1)$ is [14,16,29-31]

\begin{equation}
\left\{ i\frac{d}{dx}\left(
\begin{array}{cc}
0 & -1 \\
1 & 0%
\end{array}
\right) +\left[ E_{n}-V(x)\right] \left(
\begin{array}{cc}
0 & 1 \\
1 & 0%
\end{array}
\right) -m\left(
\begin{array}{cc}
1 & 0 \\
0 & 1%
\end{array}
\right) \right\} \psi _{nq}(x)=0,
\end{equation}
where the spinor radial wave function $\psi _{nq}(x)=\left(
\begin{array}{c}
u_{nq}(x) \\
w_{nq}(x)%
\end{array}
\right) $ is normalized with $u_{nq}(x)$ ($w_{nq}(x)$), the upper
(lower) spinor component (the $x$ range is from $-\infty $ to
$\infty $ on the full-line problem)$.$ The interaction among
nuclei is commonly described by using a potential which consists
of the Coulomb and the nuclear potentials. It is usually taken in
the form of Woods-Saxon (WS) potential. Here we take the following
Hermitian real-valued $1D$ generalized WS potential which is
specified by the shape (deformation) parameter, $q,$ [22,32,33]
\begin{equation}
V_{q}(x)=-V_{0}\frac{e^{-\alpha x}}{1+qe^{-\alpha x}},\text{ \ }\alpha =1/a,%
\text{ }r-R_{0}\rightarrow x,\text{ \ }q\geq 0,\text{ }R_{0}\gg a,
\end{equation}
where $r$ refers to the center-of-mass distance between the
projectile and the target nuclei (the $r$ range is from $0$ to
$\infty $). The relevant
parameters of the inter-nuclear potential are given as follows: $R_{\text{0}%
}=r_{0}A^{1/3}$ is to define the confinement barrier position
value of the corresponding spherical nucleus or the width of the
potential, $A$ is the
target mass number, $r_{0}$ is the radius parameter, the field strength $%
V_{0}$ controls the barrier height of the Coulombic part, $a$ is
the surface diffuseness parameter has to control its slope, which
is usually adjusted to the experimental values of ionization
energies. Note further, $q$ is a shape (deformation) parameter,
the strength of the exponential part other than unity, set to
determine the shape of potential and is arbitrarily taken to be a
real constant within the potential. It should be noted that the
spatial coordinates in the potential are not deformed and thus the
potential still remains spherical.

It is worth to state that under radial coordinate transformation, $%
r\rightarrow r+\Delta ,$ then the generalized WS potential in Eq.
(12) changes into the standard WS potential ($q=1)$ but with the
displacement parameter $\Delta $ satisfies the expression $\exp
(\Delta /a)=q$ and with a field strength $V_{0}^{\prime
}=V_{0}\exp (-\Delta /a)$ [32]. The sense of generalization or
deformation of the potential becomes clear. For completeness, it
could be stated that if $\Delta $ is positive (corresponding to
$q>1)$ then one may need to impose the condition on the choice of
$\Delta ,$ that is, $\left| \Delta \right| \ll R_{0}.$

Obviously, for some specific $q$ values this potential reduces to
the
well-known types, such as for $q=0$ to the exponential potential and for $%
q=-1$ and $a=\delta ^{-1}$ to the generalized Hulth\'{e}n
potential (cf. [20,34] and the references therein).

For the given generalized WS potential, Eq. (11) decomposes into
the following two components:

\begin{equation}
\left( i\frac{d}{dx}+E_{nq}+V_{0}\frac{e^{-\alpha x}}{1+qe^{-\alpha x}}%
\right) u_{nq}(x)=mw_{nq}(x),
\end{equation}
\begin{equation}
\left( -i\frac{d}{dx}+E_{nq}+V_{0}\frac{e^{-\alpha x}}{1+qe^{-\alpha x}}%
\right) w_{nq}(x)=mu_{nq}(x).
\end{equation}
It should also be noted that the coupled differential equations
allow finite bound states (real) solutions as $u_{nq}(r\rightarrow
\infty )\rightarrow 0$ and $w_{nq}(r\rightarrow \infty
)\rightarrow 0.$ The above set of coupled equations can be reduced
to a second-order differential equation. Hence, combining the last
two equations, this provides the following second-order
Schr\"{o}dinger-type equation for upper (lower) components,
respectively, as

\begin{equation}
u{}_{nq}^{\prime \prime }(x)+\left[ \widetilde{E}_{nq}+V_{1}\frac{%
e^{-2\alpha x}}{(1+qe^{-\alpha x})^{2}}+V_{2}\frac{e^{-\alpha x}}{%
1+qe^{-\alpha x}}\right] u_{nq}(x)=0,
\end{equation}
\begin{equation}
w{}_{nq}^{\prime \prime }(x)+\left[ \widetilde{E}_{nq}+V_{1}^{\ast }\frac{%
e^{-2\alpha x}}{(1+qe^{-\alpha x})^{2}}+V_{2}^{\ast }\frac{e^{-\alpha x}}{%
1+qe^{-\alpha x}}\right] w_{nq}(x)=0,
\end{equation}
with the following definitions

\begin{equation}
\widetilde{E}_{nq}=E_{nq}^{2}-m^{2},\text{
}V_{1}=V_{0}^{2}-iqV_{0}/a,\text{ }V_{2}=2E_{nq}V_{0}+iV_{0}/a,
\end{equation}
and $V_{1}^{\ast }$ ($V_{2}^{\ast }$) is the complex conjugate of $V_{1}$ ($%
V_{2}$), respectively.

Using the NU method, we are set to obtain bound states solutions
(relativistic energy spectrum and spinor wavefunctions) of a
spin-zero particle for a three parameter \{$V_{0},q,\alpha
$\}generalized WS
potential. We employ the following dimensionless transformation parameter, s$%
(x)=(1+qe^{-\alpha x})^{-1}$ , which maintains the transformed
wavefunctions finite, on the boundary conditions (i.e., $0\leq
r<\infty \rightarrow -\infty \leq x<\infty \rightarrow 0\leq s\leq
1)$ [21,22]$.$ Hence, Eq. (15) is reduced into the generalized
equation of hypergeometric type which is given by Eq. (1):

\begin{equation}
\frac{d^{2}u_{nq}(s)}{ds^{2}}+\frac{1-2s}{(s-s^{2})}\frac{du_{nq}(s)}{ds}+%
\frac{a^{2}}{(s-s^{2})^{2}}\left[ \widetilde{E}_{nq}+\frac{V_{1}}{q^{2}}%
(1-s)^{2}+\frac{V_{2}}{q}(1-s)\right] u_{nq}(s)=0,
\end{equation}
where we have set $u_{nq}(x)=u_{nq}(s).$ Therefore, with the
dimensionless definitions
\begin{equation}
-\epsilon ^{2}=a^{2}\widetilde{E}_{nq}\geq 0,\text{ \ \ }\beta
^{2}=a^{2}V_{2}/q,\text{ \ }\gamma ^{2}=a^{2}V_{1}/q^{2},\text{ }%
(E_{nq}^{2}\leq m^{2},\text{ }\beta ^{2}>0,\text{ }\gamma ^{2}>0),
\end{equation}
for bound states (i.e., real $\epsilon ^{2}$), one can arrive at
the simple hypergeometric equation given by

\begin{equation}
u{}_{nq}^{\prime \prime }(s)+\frac{1-2s}{(s-s^{2})}u{}_{nq}^{\prime }(s)+%
\frac{\left[ s^{2}\gamma ^{2}-s\left( \beta ^{2}+2\gamma ^{2}\right) +\text{%
\ }\beta ^{2}+\gamma ^{2}-\epsilon ^{2}\right] }{\left( s-s^{2}\right) ^{2}}%
u_{nq}(s)=0.
\end{equation}
Before further proceeding, it is necessary to compare the last
equation with Eq. (1) to obtain the following polynomials:

\begin{equation}
\widetilde{\tau }(s)=1-2s,\text{ \ \ \ }\sigma (s)=(s-s^{2}),\text{ \ \ }%
\widetilde{\sigma }(s)=s^{2}\gamma ^{2}-s\left( \beta ^{2}+2\gamma
^{2}\right) +\text{\ }\beta ^{2}+\gamma ^{2}-\epsilon ^{2}.
\end{equation}
The substitution of the above expressions into Eq. (9), together with $%
\sigma ^{\prime }(s)=1-2s,$ gives

\begin{equation}
\pi (s)=\pm i\sqrt{s^{2}(\gamma ^{2}+k)-s\left( \beta ^{2}+2\gamma
^{2}+k\right) +\text{\ }\beta ^{2}+\gamma ^{2}-\epsilon ^{2}}.
\end{equation}
It is taken into consideration that the discriminant of the square
root sign has to be zero. Hence, the second-order equation under
the expected roots
are obtained as $k_{1,2}=\beta ^{2}-2\epsilon ^{2}\pm 2\epsilon b,$ where $b=%
\sqrt{\epsilon ^{2}-\text{\ }\beta ^{2}-\gamma ^{2}}=-\left(
n+1+iaV_{0}/q\right) -\epsilon $ with $n=0,1,2,\cdots .$ In this
case, substituting these values for each $k$ into Eq. (22), the
possible solutions are obtained for $\pi (s)$ are:

\begin{equation}
\pi (s)=\pm \left\{
\begin{array}{c}
\left( b-\epsilon \right) s-b;\text{ \ \ \ for \ \ }k_{1}=\beta
^{2}-2\epsilon ^{2}+2\epsilon b, \\
\left( b+\epsilon \right) s-b;\text{ \ \ for \ \ }k_{2}=\beta
^{2}-2\epsilon
^{2}-2\epsilon b.%
\end{array}
\right.
\end{equation}
For bound state solutions, it is necessary to choose one of the
four possible forms in the last equation. Therefore, the most
suitable form is established by

\begin{equation}
\pi (s)=-\left( b+\epsilon \right) s+b,\text{ }k=\beta
^{2}-2\epsilon ^{2}-2\epsilon b.
\end{equation}
The following track in this selection is to achieve the condition
$\tau
%TCIMACRO{\U{b4}}%
%BeginExpansion
{\acute{}}%
%EndExpansion
(s)<0$ in Eq. (5), which can be obtained as
\[
\tau \text{(s)}=-2(1+b+\epsilon )s+1+2b,
\]

\begin{equation}
\tau
%TCIMACRO{\U{b4}}%
%BeginExpansion
{\acute{}}%
%EndExpansion
(s)=-2(1+b+\epsilon )=2(n+iaV_{0}/q).
\end{equation}
A particular solution can be calculated by using Eqs. (6) and
(10). Consequently, this solution is obtained as
\[
\lambda =-\gamma ^{2}-(b+\epsilon )(b+\epsilon +1),
\]

\begin{equation}
\lambda _{n}=n^{2}+n+2n\left( \epsilon +b\right) .
\end{equation}
After setting $\lambda _{n}=\lambda $ and solving for $E_{nq},$ we
find the Dirac exact energy spectra as

\[
E_{nq}=-\left[ \frac{V_{0}}{2q}\pm \kappa _{n}(q,a,V_{0})\sqrt{\frac{m^{2}}{%
V_{0}^{2}+\kappa _{n}^{2}(q,\alpha
,V_{0})}-\frac{1}{4q^{2}}}\right] ,
\]
\begin{equation}
\kappa _{n}(q,\alpha ,V_{0})=-\left[ iV_{0}+q\alpha (n+1)\right] ,\text{ }%
n=0,1,2,\cdots
\end{equation}
We should point out that E\u{g}rifes and Sever [16] have recently
obtained a similar expression to Eq. (27) for the case of the
Hulth\'{e}n potential. Let us now find the corresponding
wavefunctions. We have seen that the energy expression has a
complex form for the potential under study. We look for the
complex generalized WS potential forms that have real spectrum.

Moreover, the restriction which gives the critical coupling value
leads to the result
\begin{equation}
n\leq \frac{1}{q\alpha }\left(
\sqrt{4q^{2}m^{2}-V_{0}^{2}}-iV_{0}\right) -1,
\end{equation}
that is, there are only finitely many eigenvalues. In order that
at least one level might exist, it is necessary that the
inequality
\begin{equation}
q\alpha +iV_{0}\preceq \sqrt{4q^{2}m^{2}-V_{0}^{2}},
\end{equation}
is fulfilled. As can be seen from Eq. (28), there are only two
lower-lying states for the Dirac particle of mass unity when the
parameters $\alpha =1,$
$q=\pm 1$ for any given $V_{0}:$%
\begin{equation}
n\leq \pm \left( \sqrt{4-V_{0}^{2}}-iV_{0}\right) -1.
\end{equation}
(i) Choosing $q=1,$ the potential form (12) is reduced to the
shifted WS potential:
\begin{equation}
V(x)=-V_{0}+\frac{V_{0}}{1+e^{-\alpha x}},
\end{equation}
and then it's energy spectra yield
\begin{equation}
E_{n}=-\frac{V_{0}}{2}\pm \left[ iV_{0}+\alpha (n+1)\right] \sqrt{\frac{m^{2}%
}{V_{0}^{2}+\left[ iV_{0}+\alpha (n+1)\right] ^{2}}-\frac{1}{4}},\text{ }%
n=0,1,2,\cdots
\end{equation}
(ii) Choosing $q=-1,$ the potential form (12) is reduced to the shifted Hulth%
\'{e}n potential:
\begin{equation}
V(x)=V_{0}-\frac{V_{0}}{1-e^{-\alpha x}},
\end{equation}
and then the resulting energy eigenvalues become
\begin{equation}
E_{n}=\frac{V_{0}}{2}\pm \left[ iV_{0}-\alpha (n+1)\right] \sqrt{\frac{m^{2}%
}{V_{0}^{2}+\left[ iV_{0}-\alpha (n+1)\right] ^{2}}-\frac{1}{4}.},\text{ }%
n=0,1,2,\cdots
\end{equation}
(iii) For the case $q\rightarrow 0,$ the potential expression (12)
is reduced to the exponential potential:
\begin{equation}
V(x)=-V_{0}e^{-\alpha x},
\end{equation}
the eigenvalues expression (27) does not give an explicit form,
i.e., the NU method is not applicable to the exponential potential
(34).

Note that for this potential there is no explicit form of the
energy expression of bound states for Schr\"{o}dinger [18], KG
[35] and also Dirac [17] equations.

In addition, it can be seen easily that while the field strength $%
V_{0}\rightarrow 0,$ the energy eigenvalues yield:
\begin{equation}
E_{n}=\pm \frac{1}{2}\sqrt{4m^{2}-(n+1)^{2}\alpha ^{2}},\text{ }%
n=0,1,2,\cdots
\end{equation}
Note that in the above equation there exist bound states for the
ground and excited states $(n=0,1)$ which are $E_{0}=\pm
\sqrt{3}m/2$ and $E_{1}=0,$ respectively, for positive $q$ values
and $a=\lambda _{c},$ where $\lambda _{c}=1/m$ denotes the Compton
wavelength of the Dirac particle. Otherwise, there are no bound
states for $n\geq 2$ states.

On the other hand, for the same value of $\alpha $ and negative
$q$ values when $V_{0}\rightarrow 0,$ all energy eigenvalues go to
zero. If the value of $q$ is increasing, all positive bound states
go to zero, from (27), asymptotically.

An inspection of the energy expression given by Eq. (27), for any given $%
\alpha ,$ shows that we deal with a family of generalized WS
potentials. The sign of $V_{0}$ does not effect the bound states.
The spectrum consists of complex eigenvalues depending on $q.$ As
we shall see the role played by the range parameter $\alpha $ is
very crucial in this regard. Of course, it is clear that by
imposing appropriate changes in the parameters $\left\{ \alpha
,V_{0},q\right\} ,$ the energy spectrum in Eq. (27) for any
modified parameter can be also calculated by resolving Dirac
equation for every parameter change.

Let us calculate the wavefunctions. Inserting, $\pi (s)$ and
$\sigma (s)$ in Eq. (4) and consequently solving the resulting
first-order differential equation, we find

\begin{equation}
\phi _{n}(s)=s^{b}(1-s)^{\epsilon }.
\end{equation}
In addition, to find the function, $y_{nq}(s),$ which is the
polynomial solution of hypergeometric-type equation, we multiply
Eq. (3) by $\rho (s)$ so that it can be written in self-adjoint
form [28]

\begin{equation}
(\sigma (s)\rho (s)y_{nq}^{\prime }(s))^{\prime }+\lambda \rho
(s)y_{nq}(s)=0,
\end{equation}
where $\rho (s)$ satisfies Eq. (8), which gives

\begin{equation}
\rho (s)=s^{2b}(1-s)^{2\epsilon }.
\end{equation}
The second eigenfunction can be obtained by Eq. (7) as

\begin{equation}
y_{nq}(s)=D_{nq}s^{-2b}(1-s)^{-2\epsilon
}\frac{d^{n}}{ds^{n}}\left[ s^{n+2b}\left( 1-s\right)
^{n+2\epsilon }\right] ,
\end{equation}
where $D_{nq}$ is a normalization constant$.$ In the limit
$q\rightarrow 1,$ the polynomial solutions of $\ y_{n}(s)$ are
expressed in terms of Jacobi Polynomials, which is one of the
classical orthogonal polynomials, with
weight function given by Eq. (39) for $s\in $ $\left[ 0,1\right] ,$ giving $%
y_{n}(s)\simeq P_{n}^{(2b,2\epsilon )}(1-2s).$ Obviously, the
radial wave function $u_{nq}(s)$ for the ${\rm s}$-wave can be
obtained by substituting Eqs. (37) and (40) into Eq. (2) as

\begin{equation}
u_{n}(s)=N_{nq}s^{b}(1-s)^{\epsilon }P_{n}^{(2b,2\epsilon
)}(1-2s),
\end{equation}
where $s(r)=(1+e^{-(r-R_{0})/a})^{-1},$ $R_{0}\gg a$ and $N_{nq}$
is a new normalization constant. Further, by using the
differential and recursion properties of the Jacobi polynomials
[36], the lower spinor component can be obtained from Eq. (13) as

\[
mw_{nq}(s)=N_{nq}s^{b}(1-s)^{\epsilon }\left\{ E_{nq}-i\alpha
\epsilon + \left[ \frac{V_{0}}{q}-i\alpha \left(
n+1+\frac{iV_{0}}{\alpha q}\right) (1-s)\right] \right\}
P_{n}^{(2b,2\epsilon )}(1-2s)
\]

\begin{equation}
+N_{nq}i\alpha \left( n+1+\frac{2iV_{0}}{q\alpha }\right)
s^{b}(1-s)^{\epsilon }P_{n}^{(2b+1,2\epsilon +1)}(1-2s).
\end{equation}
Notice the well behavior of the wave function at infinity. As an
example, the ground state wave function behaves like

\begin{equation}
u_{0}(s\rightarrow 0)\rightarrow 0,\text{ }u_{0}(s\rightarrow
1)\rightarrow 0,
\end{equation}
and thus representing a truly bound-state solution. Further, we
make use of the fact that the Jacobi polynomials can be explicitly
written in two different ways [36]:

\begin{equation}
P_{n}^{(\rho ,\nu )}(z)=2^{-n}\sum\limits_{p=0}^{n}(-1)^{n-p}%
%TCIMACRO{\binom{n+\rho }{p}}%
%BeginExpansion
{n+\rho  \choose p}%
%EndExpansion
%TCIMACRO{\binom{n+\nu }{n-p}}%
%BeginExpansion
{n+\nu  \choose n-p}%
%EndExpansion
\left( 1-z\right) ^{n-p}\left( 1+z\right) ^{p},
\end{equation}

\begin{equation}
P_{n}^{(\rho ,\nu )}(z)=\frac{\Gamma (n+\rho +1)}{n!\Gamma (n+\rho +\nu +1)}%
\sum\limits_{r=0}^{n}%
%TCIMACRO{\binom{n}{r}}%
%BeginExpansion
{n \choose r}%
%EndExpansion
\frac{\Gamma (n+\rho +\nu +r+1)}{\Gamma (r+\rho +1)}\left( \frac{z-1}{2}%
\right) ^{r},
\end{equation}
where $%
%TCIMACRO{\binom{n}{r}}%
%BeginExpansion
{n \choose r}%
%EndExpansion
=\frac{n!}{r!(n-r)!}=\frac{\Gamma (n+1)}{\Gamma (r+1)\Gamma
(n-r+1)}.$ Using Eqs. (44) and (45), we obtain the following two
explicit expressions:
\[
P_{n}^{(2b,2\epsilon )}(1-2s)=(-1)^{n}\Gamma (n+2b+1)\Gamma
(n+2\epsilon +1)
\]

\begin{equation}
\times \sum\limits_{p=0}^{n}\frac{(-1)^{p}q^{n-p}}{p!(n-p)!\Gamma
(p+2\epsilon +1)\Gamma (n+2b-p+1)}s^{n-p}(1-s)^{p},
\end{equation}

\begin{equation}
P_{n}^{(2b,2\epsilon )}(1-2s)=\frac{\Gamma (n+2b+1)}{\Gamma
(n+2b+2\epsilon
+1)}\sum\limits_{r=0}^{n}\frac{(-1)^{r}q^{r}\Gamma (n+2b+2\epsilon +r+1)}{%
r!(n-r)!\Gamma (2b+r+1)}s^{r}.
\end{equation}

\[
1=N_{nq}^{2}(-1)^{n}\frac{\Gamma (n+2\epsilon +1)\Gamma
(n+2b+1)^{2}}{\Gamma
(n+2\epsilon +2b+1)}\left\{ \sum\limits_{p=0}^{n}\frac{(-1)^{p}q^{n-p}}{%
p!(n-p)!\Gamma (p+2\epsilon +1)\Gamma (n+2b-p+1)}\right\}
\]

\begin{equation}
\times \left\{ \sum\limits_{r=0}^{n}\frac{(-1)^{r}q^{r}\Gamma
(n+2\epsilon +2b+r+1)}{r!(n-r)!\Gamma (2b+r+1)}\right\}
I_{nq}(p,r),
\end{equation}
where

\begin{equation}
I_{nq}(p,r)=\int\limits_{0}^{1}s^{n+2b+r-p}(1-s)^{p+2\epsilon }ds.
\end{equation}
Using the following integral representation of the hypergeometric
function [36]

\[
\int\limits_{0}^{1}s^{\alpha _{0}-1}(1-s)^{\gamma _{0}-\alpha
_{0}-1}(1-qs)^{-\beta _{0}}ds=_{2}F_{1}(\alpha _{0},\beta _{0}:\gamma _{0};q)%
\frac{\Gamma (\alpha _{0})\Gamma (\gamma _{0}-\alpha _{0})}{\Gamma
(\gamma _{0})},
\]

\begin{equation}
\lbrack
%TCIMACRO{\func{Re}}%
%BeginExpansion
\mathop{\rm Re}%
%EndExpansion
(\gamma _{0})>%
%TCIMACRO{\func{Re}}%
%BeginExpansion
\mathop{\rm Re}%
%EndExpansion
(\alpha _{0})>0,\text{ \ }\left\vert \arg (1-q)\right\vert <\pi ],
\end{equation}%
which gives

\begin{equation}
_{2}F_{1}(\alpha _{0},\beta _{0}:\alpha _{0}+1;q)/\alpha
_{0}=\int\limits_{0}^{1}s^{\alpha _{0}-1}(1-qs)^{-\beta _{0}}ds,
\end{equation}

\[
_{2}F_{1}(\alpha _{0},\beta _{0}:\gamma _{0};q)=\frac{\Gamma
(\gamma _{0})\Gamma (\gamma _{0}-\alpha _{0}-\beta _{0})}{\Gamma
(\gamma _{0}-\alpha _{0})\Gamma (\gamma _{0}-\beta _{0})},
\]
\begin{equation}
\lbrack
%TCIMACRO{\func{Re}}%
%BeginExpansion
\mathop{\rm Re}%
%EndExpansion
(\gamma _{0}-\alpha _{0}-\beta _{0})>0,\text{ }%
%TCIMACRO{\func{Re}}%
%BeginExpansion
\mathop{\rm Re}%
%EndExpansion
(\gamma _{0})>%
%TCIMACRO{\func{Re}}%
%BeginExpansion
\mathop{\rm Re}%
%EndExpansion
(\beta _{0})>0,
\end{equation}
for $q=1.$ Setting $\alpha _{0}=n+2b+r-p+1,$ $\beta
_{0}=-p-2\epsilon ,$ and $\gamma _{0}=\alpha _{0}+1,$ one gets

\begin{equation}
I_{nq}(p,r)=\frac{_{2}F_{1}(\alpha _{0},\beta _{0}:\gamma
_{0};q)}{\alpha
_{0}}=\frac{(n+2b+r-p+1)!(p+2\epsilon )!}{(n+2b+r-p+1)(n+2\epsilon +r+2b+1)!}%
.
\end{equation}
In view of the above complex energy spectra (27), it will be of
interest to see how complex potential form of Eq. (12) would
effect this result. Therefore, we shall change the values of the
potential parameters $(\alpha ,q,V_{0})$ from real to pure
imaginary (complex) to obtain Hamiltonians that may or may not be
${\cal PT}{\rm -}$symmetric. Hence, we also make the same
parameter change in the energy spectra (27) and the upper and
lower spinor components of the wavefunction Eqs. (41) and (42)
respectively. The resulting non-Hermitian complex potential could
have real energy spectra. To
this end we consider the complexified forms of the generalized WS potential.

\subsection{Non-Hermitian ${\cal PT}{\rm -s}$ymmetric new trigonometric and
periodic potential form}

Let us consider the case where at least one of the potential
parameters be
complex. In this case, $\alpha $ is taken to be a complex parameter (i.e., $%
\alpha \rightarrow i\alpha $). Consequently, the potential in Eq.
(12) transforms into the form

\begin{equation}
V_{q}(x)=-\frac{V_{0}}{q^{2}+2q\cos (\alpha x)+1}\left[ q+\cos
(\alpha x)-i\sin (\alpha x)\right] =V_{q}^{\ast }(-x),\text{ }
\end{equation}
which is a ${\cal PT}{\rm -}$symmetric but non-Hermitian. We note
that the transformed potential in Eq. (54) has a trigonometric and
periodic form. Obviously, the last form of potential forms have no
physical relation what so ever with the WS potential in (12), any
of it's generalizations, it's well-known behavior or properties.
As we have seen, simple mathemetical manipulations have not to
cloud our intutive judgement and conceptual physical
understanding. From a mathematical prospective, it might be
possible that one can use calculus to think of the hyperbolic or
exponential functions as another form of trigonometric ones.
However, physically these functions, if considered as potentials,
are dramatically different. We should not refer to either
potential in Eq. (54) as WS-type. Nevertheless,
this new complex potential embodies their periodic, trigonometric and ${\cal %
PT}{\rm -}$symmetric behaviors. Hence, this type of potentials
(54) has real spectrum given by

\begin{equation}
E_{nq}=-\frac{V_{0}}{2q}\pm \left( V_{0}+q(n+1)/a\right) \sqrt{\frac{1}{%
4q^{2}}-\frac{m^{2}}{V_{0}^{2}-\left( V_{0}+q(n+1)/a\right)
^{2}}},
\end{equation}
if the following restriction$\ 4q^{2}m^{2}\leq V_{0}^{2}-\left(
V_{0}+q(n+1)/a\right) ^{2}$ being achieved$.$ The critical
coupling value is

\[
V_{0}\leq -\frac{q(n+1)}{2a}-\frac{2qam^{2}}{n+1},
\]
which leads to the following condition:
\[
\frac{V_{0}+q\alpha -\sqrt{V_{0}^{2}-4q^{2}m^{2}}}{q\alpha }\leq n\leq \frac{%
V_{0}+q\alpha +\sqrt{V_{0}^{2}-4q^{2}m^{2}}}{q\alpha },
\]
meaning that the number of real eigenvalues are finite. Further,
the corresponding radial wave function $u_{nq}(s)$ for the ${\rm
s}$-wave could be determined as

\begin{equation}
u_{nq}(s)=N_{nq}s^{c}(1-s)^{i\epsilon }P_{n}^{(2c,2i\epsilon
)}(1-2s),
\end{equation}
\[
mw_{nq}(s)=N_{nq}s^{c}(1-s)^{i\epsilon }\left\{ E_{nq}+i\alpha
\epsilon +
\left[ \frac{V_{0}}{q}+\alpha \left( n+1+\frac{V_{0}}{\alpha q}\right) (1-s)%
\right] \right\} P_{n}^{(2c,2i\epsilon )}(1-2s)
\]

\begin{equation}
-N_{nq}\alpha \left( n+1+\frac{2V_{0}}{q\alpha }\right)
s^{c}(1-s)^{i\epsilon }P_{n}^{(2c+1,2i\epsilon +1)}(1-2s),
\end{equation}
where $c=-(n+1+aV_{0}/q)-i\epsilon ,$ and
$s(r)=(1+qe^{-i(r-R_{0})/a})^{-1}.$

For the sake of comparing the relativistic and non-relativistic
binding energies, we need to solve the $1D$ \ Schr\"{o}dinger
equation for the complex form of the generalized WS potential
given by Eq. (54). Employ a
convenient transformation given by $s(r)=(1+qe^{-i(r-R_{0})/a})^{-1}$, $%
0\leq r\leq \infty \rightarrow 0\leq s(r)\leq 1,$ we obtain [21]

\begin{equation}
\psi {}_{nq}^{\prime \prime }(s)+\frac{1-2s}{(s-s^{2})}\psi
{}_{nq}^{\prime }(s)+\frac{\left[ -\beta ^{2}s+\beta ^{2}-\epsilon
^{2}\right] }{\left( s-s^{2}\right) ^{2}}\psi _{nq}(s)=0,
\end{equation}
for which
\[
\widetilde{\tau }(s)=1-2s,\text{ \ \ \ }\sigma (s)=s-s^{2},\text{ \ \ }%
\widetilde{\sigma }(s)=-\beta ^{2}s+\beta ^{2}-\epsilon ^{2},
\]

\begin{equation}
\epsilon ^{2}=\frac{2ma^{2}}{\hbar ^{2}}E_{nq},\text{ \ }\beta ^{2}=-\frac{%
2ma^{2}}{\hbar ^{2}q}V_{0}\text{ \ \ (}E_{nq}<0,\text{ }\beta
^{2}>0).
\end{equation}
The function $\tau (s)$ could be obtained as

\begin{equation}
\tau (s)=-2(1+d+\epsilon )s+(1+2d),\text{ }d=\sqrt{\epsilon ^{2}-\text{\ }%
\beta ^{2}},
\end{equation}
if $\pi (s)=-(d+\epsilon )s+d$ is chosen for $k_{-}=-(d+\epsilon
)^{2}.$ We can also find the eigenvalues from Eqs. (6) and (10) as
\begin{equation}
\lambda =-\left( d+\epsilon \right) (d+\epsilon +1),\text{ \
}\lambda _{n}=2n\left( d+\epsilon +1\right) +n(n-1).
\end{equation}
Thus, solving Eq. (61) for the energy eigenvalues, we obtain
\begin{equation}
E_{nq}(V_{0},i\alpha )=\frac{\hbar ^{2}}{2ma^{2}}\left[ \frac{n+1}{2}-\frac{%
\gamma }{(n+1)}\right] ^{2},\text{ \ \ }\gamma
=\frac{ma^{2}V_{0}}{\hbar ^{2}q},\text{ \ }0\leq n<\infty .\text{
\ \ }
\end{equation}
On the other hand, the radial wavefunctions in the present case
become

\begin{equation}
\psi _{nq}(s)=N_{nq}s^{d}(1-s)^{\epsilon }P_{n}^{(2d,2\epsilon
)}(1-2s),
\end{equation}
with $s(r)=(1+qe^{-i(r-R_{0})/a})^{-1}$ and $N_{nq}$ is a new
normalization constant determine by

\[
1=N_{nq}^{2}(-1)^{n}\frac{(n+2\epsilon )!\Gamma
(n+2d+1)^{2}}{\Gamma
(n+2d+2\epsilon +1)}\left\{ \sum\limits_{p=0}^{n}\frac{(-1)^{p}q^{n-p}}{%
p!(n-p)!(2\epsilon +p)!\Gamma (n+2d-p+1)}\right\}
\]

\begin{equation}
\times \left\{ \sum\limits_{r=0}^{n}\frac{(-1)^{r}q^{r}\Gamma
(n+2d+r+2\epsilon +1)}{r!(n-r)!\Gamma (2d+r+1)}\right\}
I_{nq}(p,r),
\end{equation}
where the integral $I_{nq}(p,r)=\int\limits_{0}^{1}s^{n+2d+r-p}(1-qs)^{p+2%
\epsilon }ds$ is given by

\begin{equation}
I_{nq}(p,r)=_{2}F_{1}(n+2c+r-p+1,-p-2\epsilon
:n+2c+r-p+2:1)B(n+2c+r-p+1,1),
\end{equation}
Figures 1 and 2 show the variation of the ground-state (i.e.,
$n=0$) as a function of the coupling constant $V_{0}$ for
different positive and negative $q$, and $a=\lambda _{c}.$
Obviously, in Figure 1, the non-Hermitian ${\cal PT}$-symmetric
generalized WS potential generates real and negative bound-states
for $q>0,$ it generates real and positive bound-states for the
same value of $\alpha $ when $q<0$ (Figure 2). Further, Figures 3
and 4 show the variation of the first three energy eigenstates as
a function of $\alpha $ for (a) $q=1.0$ and (b) $q=-1.0$ with
$V_{0}=2.5m.$ Obviously, for the given $V_{0},$ as seen from
Figures 3 and 4 all possible eigenstates have negative (positive)
eigenenergies if the parameter $q$ is positive (negative). It is
almost notable that there are some crossing points of the
relativistic energy eigenvalues for some $V_{0}$ values.

\subsection{Non-Hermitian non-${\cal PT}{\rm -s}$ymmetric generalized
Woods-Saxon potential}

In this part, we consider two parameters $\left\{ V_{0},q\right\}
$ to be complex parameters (i.e., $V_{0}\rightarrow iV_{0},$
$q\rightarrow iq$). Consequently, the potential in Eq. (12)
transforms to the following form

\begin{equation}
V_{q}(x)=V_{0}\frac{\left[ 2\cosh ^{2}(\alpha x)-\sinh (2\alpha
x)-1\right] -i\left[ \cosh (\alpha x)-\sinh (\alpha x)\right]
}{1+q^{2}\left[ 2\cosh ^{2}(\alpha x)-\sinh (2\alpha x)-1\right]
},
\end{equation}
which is a non-${\cal PT}{\rm -}$symmetric but non-Hermitian. The
complex energy eigenvalues of Eq. (66) are given by

\begin{equation}
E_{nq}=-\frac{V_{0}}{2q}\pm i\left( iV_{0}+q(n+1)/a\right) \sqrt{\frac{1}{%
4q^{2}}-\frac{m^{2}}{V_{0}^{2}+\left( iV_{0}+q(n+1)/a\right)
^{2}}},
\end{equation}
On the other hand, the corresponding radial wave functions
$u_{nq}(s)$ for the ${\rm s}$-wave could be determined as

\begin{equation}
u_{nq}(s)=N_{nq}s^{b}(1-s)^{\epsilon }P_{n}^{(2b,2\epsilon
)}(1-2s),
\end{equation}
and the lower spinor component $w_{nq}(s)$ is given by Eq. (42) with $%
s(r)=(1+iqe^{-(r-R_{0})/a})^{-1}.$ The integral $I_{nq}(p,r)=\int%
\limits_{0}^{1}s^{n+2b+r-p}(1-s)^{p+2\epsilon }ds$ is given by

\begin{equation}
I_{nq}(p,r)=_{2}F_{1}(n+2b+r-p+1,-p-2\epsilon
:n+2b+r-p+2;i)B(n+2b+r-p+1,1).
\end{equation}

\subsection{Pseudo-Hermiticity and ${\cal PT}{\rm -s}$ymmetric new
trigonometric and periodic potential form}

Finally, in this part, when all the parameters $\left\{
V_{0},\alpha
,q\right\} $ are complex parameters (i.e., $V_{0}\rightarrow iV_{0},$ $%
\alpha \rightarrow i\alpha ,$ $q\rightarrow iq$), from Eq. (12) we
obtain

\begin{equation}
V_{q}(x)=-\frac{V_{0}}{q^{2}+2q\sin (\alpha x)+1}\left[ q+\sin
(\alpha x)+i\cos (\alpha x)\right] =V_{q}^{\ast }(\frac{\pi
}{\alpha }-x).
\end{equation}
Note that the transformed potential in Eq. (70) has a
trigonometric and periodic form. As we remarked earlier, the above
potential forms do not have any of the WS potential properties.
Nonetheless, they have periodic, trigonometric and ${\cal PT}{\rm
-}$symmetric behaviors. The potential in Eq. (70) is a
pseudo-Hermitian potential [37,38] having a $\pi /\alpha $
phase difference with respect to the potential (I), it is also a ${\cal PT}%
{\rm -}$symmetric, $\eta =P$-pseudo-Hermitian (i.e., $%
PTV_{q}(x)(PT)^{-1}=V_{q}(x),$ with $P=\eta :x\rightarrow \frac{\pi }{%
2\alpha }-x$ and $T:i\rightarrow -i)$ but non-Hermitian having
real spectrum given by

\begin{equation}
E_{nq}=-\frac{V_{0}}{2q}\mp \left( V_{0}+q(n+1)/a\right) \sqrt{\frac{1}{%
4q^{2}}-\frac{m^{2}}{V_{0}^{2}-\left( V_{0}+q(n+1)/a\right)
^{2}}},
\end{equation}
if the same restrictions after Eq. (55) are achieved.

On the other hand, the corresponding radial wave functions
$u_{nq}(s)$ for the ${\rm s}$-wave could be determined as

\begin{equation}
\psi _{nq}(s)=N_{nq}s^{c}(1-s)^{i\epsilon }P_{n}^{(2c,2i\epsilon
)}(1-2s),
\end{equation}
and the lower spinor component $w_{nq}(s)$ is given by Eq. (48) with $%
s(r)=(1+iqe^{-i(r-R_{0})/a})^{-1}.$ The integral $I_{nq}(p,r)=\int%
\limits_{0}^{1}s^{n+2c+r-p}(1-s)^{p+2i\epsilon }ds$ is given by

\begin{equation}
I_{nq}(p,r)=_{2}F_{1}(n+2c+r-p+1,-p-2i\epsilon
:n+2c+r-p+2;i)B(n+2c+r-p+1,1).
\end{equation}

\section{The Solution of the Generalized WS Potential for $q=0$}

\label{TS}We have obtained the bound state solutions of the
generalized WS potential with $q\neq 0$ and the explicit form of
the eigenvalues and the spinor wavefunctions. In addition, we
start finding solutions for $q=0$ case with the definition of new
variable $s=e^{-\alpha x}.$ Hence, Eq. (18) is reduced to the
generalized equation of hypergeometric type:
\begin{equation}
\frac{d^{2}u_{n}(s)}{ds^{2}}+\frac{1}{s}\frac{du_{n}(s)}{ds}+\frac{1}{s^{2}}%
\left[ -\epsilon ^{2}+\text{\ }\beta s+\gamma s^{2}\right]
u_{n}(s)=0.
\end{equation}
Further, we define the following dimensionless expressions:
\begin{equation}
\epsilon ^{2}=-\frac{1}{\alpha ^{2}}(E^{2}-m^{2}),\text{ \ \ }\beta =i\frac{%
V_{0}}{\alpha }+\frac{2EV_{0}}{\alpha ^{2}},\text{ \ }\gamma =\frac{V_{0}^{2}%
}{\alpha ^{2}},
\end{equation}

\begin{equation}
\widetilde{\tau }(s)=1,\text{ }\sigma (s)=s,\text{ \ }\widetilde{\sigma }%
(s)=-\epsilon ^{2}+\text{\ }\beta s+\gamma s^{2},
\end{equation}
with real $\epsilon ^{2}>0$ $(E^{2}<m^{2})$ for bound states. The
substitution of the above expressions into Eq. (9), together with
$\sigma ^{\prime }(s)=1,$ gives

\begin{equation}
\pi (s)=\pm i\sqrt{\gamma s^{2}-(k-\beta )s-\epsilon ^{2}}.
\end{equation}
Substituting the value for each $k$ into the last equation, we
obtain

\begin{equation}
\pi (s)=\pm \left\{
\begin{array}{c}
i\delta s+\epsilon ;\text{ \ \ \ for \ \ }k=\beta +2i\delta \epsilon , \\
i\delta s-\epsilon ;\text{ \ \ for \ \ }k=\beta -2i\delta \epsilon ,%
\end{array}
\right.
\end{equation}
where $\delta =\sqrt{\gamma }=\frac{V_{0}}{\alpha }.$ Therefore,
the most suitable form is established by

\begin{equation}
\pi (s)=-i\delta s+\epsilon ,\text{ }k=\beta -2i\delta \epsilon .
\end{equation}
The following track in this selection is to achieve the condition
$\tau
%TCIMACRO{\U{b4}}%
%BeginExpansion
{\acute{}}%
%EndExpansion
(s)<0$ in Eq. (5), which can be obtained as
\begin{equation}
\tau \text{(s)}=-2i\delta s+1+2\epsilon ,\text{ }\tau
%TCIMACRO{\U{b4}}%
%BeginExpansion
{\acute{}}%
%EndExpansion
(s)=-2i\delta ,\text{ }\lambda =\beta -2i\delta \epsilon -i\delta
.
\end{equation}
Let us calculate the wavefunctions. Inserting, $\pi (s)$ and
$\sigma (s)$ in Eq. (4) and consequently solving the resulting
first-order differential equation, we find

\begin{equation}
\phi _{n}(s)=s^{\epsilon }e^{-iV_{0}s/\alpha }.
\end{equation}
In addition, to find the function, $y_{nq}(s),$ which is the
polynomial solution of hypergeometric-type equation, we multiply
Eq. (3) by $\rho (s)$ so that it can be written in self-adjoint
form [28]

\begin{equation}
sy_{n}^{\prime \prime }+\left[ 1+2\epsilon -\nu s\right] y_{n}^{\prime }-%
\left[ \left( \epsilon +\frac{1}{2}\right) \nu -\beta \right]
y_{n}=0,\text{ }\nu =2i\delta .
\end{equation}
The solution can be written in terms of confluent hypergeometric
function as follows:

\begin{equation}
y_{n}(s)=_{1}F_{1}\left( \epsilon +\frac{1}{2}+i\frac{\beta \alpha }{2V_{0}}%
,1+2\epsilon ,\frac{2iV_{0}}{\alpha }s\right) ,
\end{equation}
and consequently the upper spinor

\begin{equation}
u(s)=A_{1}F_{1}\left( \epsilon +\frac{1}{2}+i\frac{\beta \alpha }{2V_{0}}%
,1+2\epsilon ,\frac{2iV_{0}}{\alpha }s\right) s^{\epsilon
}e^{-iV_{0}s/\alpha }.
\end{equation}
Finally, the lower spinor is found by

\begin{equation}
mw(s)=\left[ i\frac{d}{dx}+E+V_{0}s\right] A_{1}F_{1}\left( \epsilon +\frac{1%
}{2}+i\frac{\beta \alpha }{2V_{0}},1+2\epsilon ,\frac{2iV_{0}}{\alpha }%
s\right) s^{\epsilon }e^{-iV_{0}s/\alpha }.
\end{equation}

\section{Results and Conclusions}
In this work, we have seen that the $s$-wave Dirac equation with
the generalized WS potential can be solved exactly for its bound
states using the hypergeometric method. The relativistic bound
states energy spectrum and the corresponding wave functions for
the generalized WS potential have been obtained by the
hypergeometric method. Some interesting
results including the ${\cal PT}{\rm -}$symmetric, non-${\cal PT}{\rm -}$%
symmetric non-Hermitian, and non-${\cal PT}{\rm -}$symmetric $P$%
-pseudo-Hermitian versions of the generalized WS potential have
also been discussed for bound states. In addition, we have
discussed the relation between the non-relativistic and
relativistic solutions and the possibility of existence of bound
states for complex parameters.

\section{Acknowledgments}
This research was partially supported by the Scientific and
Technical Research Council of Turkey.

\newpage

\begin{figure}
\caption{The ground-state $(n=0)$ energy, in a non-Hermitian ${\cal PT}$%
-symmetric potential given by Eq. (54), as a function of the
coupling constant $V_{0}$ for three positive shape parameters $q$ with $a=\protect%
\lambda _{c}.$}
\end{figure}

\begin{figure}
\caption{The ground-state $(n=0)$ energy, in a non-Hermitian ${\cal PT}$%
-symmetric potential given by Eq. (54), as a function of the
coupling constant $V_{0}$ for three negative shape parameters $q$ with $a=\protect%
\lambda _{c}.$}
\end{figure}

\begin{figure}
\caption{The first three energy eigenstates, in a non-Hermitian ${\cal PT}$%
-symmetric potential given by Eq. (54), as a function of the range
parameter $\protect\alpha $ for a negative shape parameter
$(q=-1.0)$ with a coupling constant $V_{0}=2.5m.$}
\end{figure}

\begin{figure}
\caption{The first three energy eigenstates, in a non-Hermitian ${\cal PT}$%
-symmetric potential given by Eq. (54), as a function of the range
parameter $\protect\alpha $ for a positive shape parameter
$(q=1.0)$ with a coupling constant $V_{0}=2.5m.$}
\end{figure}

\newpage

\newpage

\begin{figure}[htbp]
\centering
\includegraphics[height=5in, width=7in]{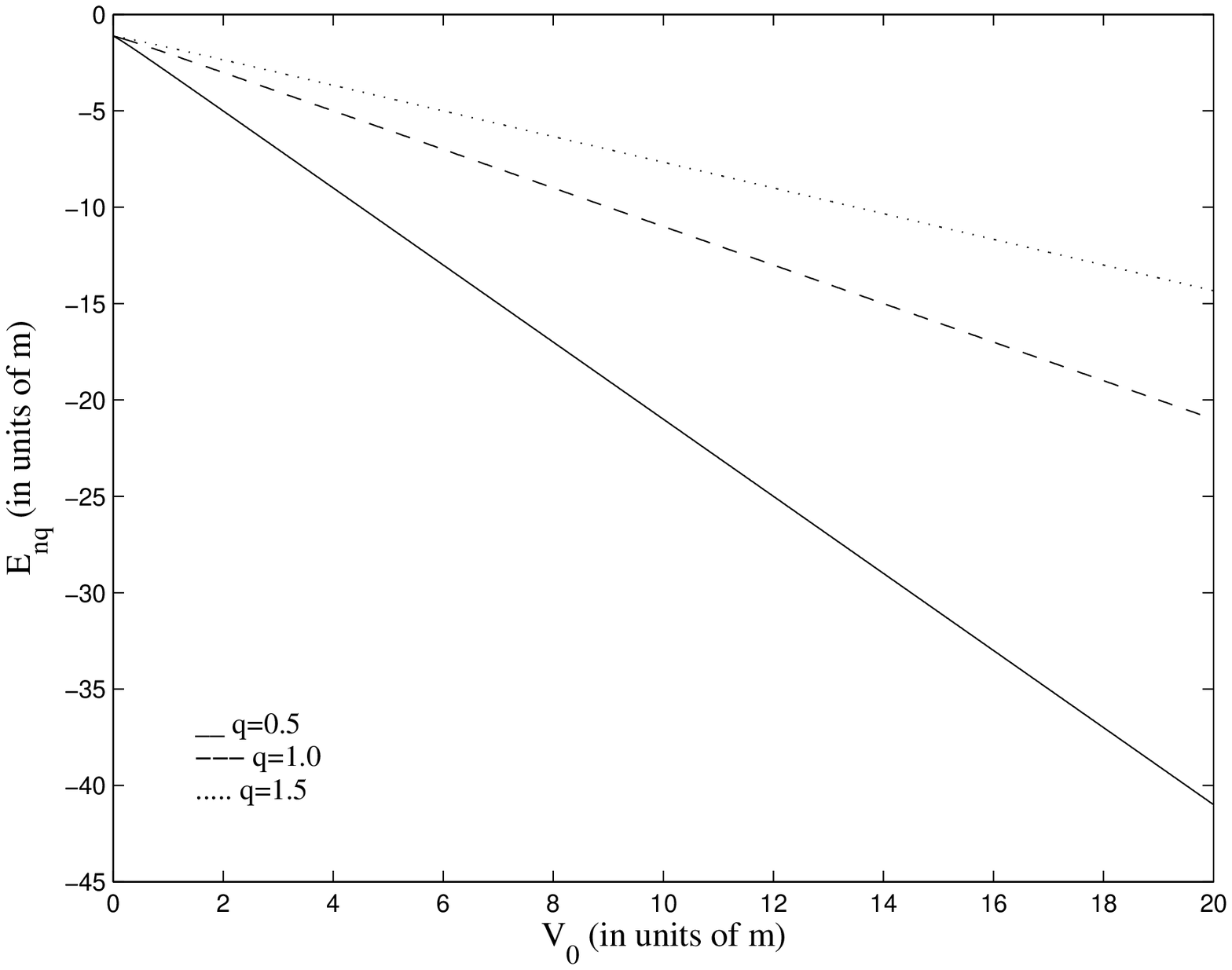}
\end{figure}

\begin{figure}[htbp]
\centering
\includegraphics[height=5in, width=7in]{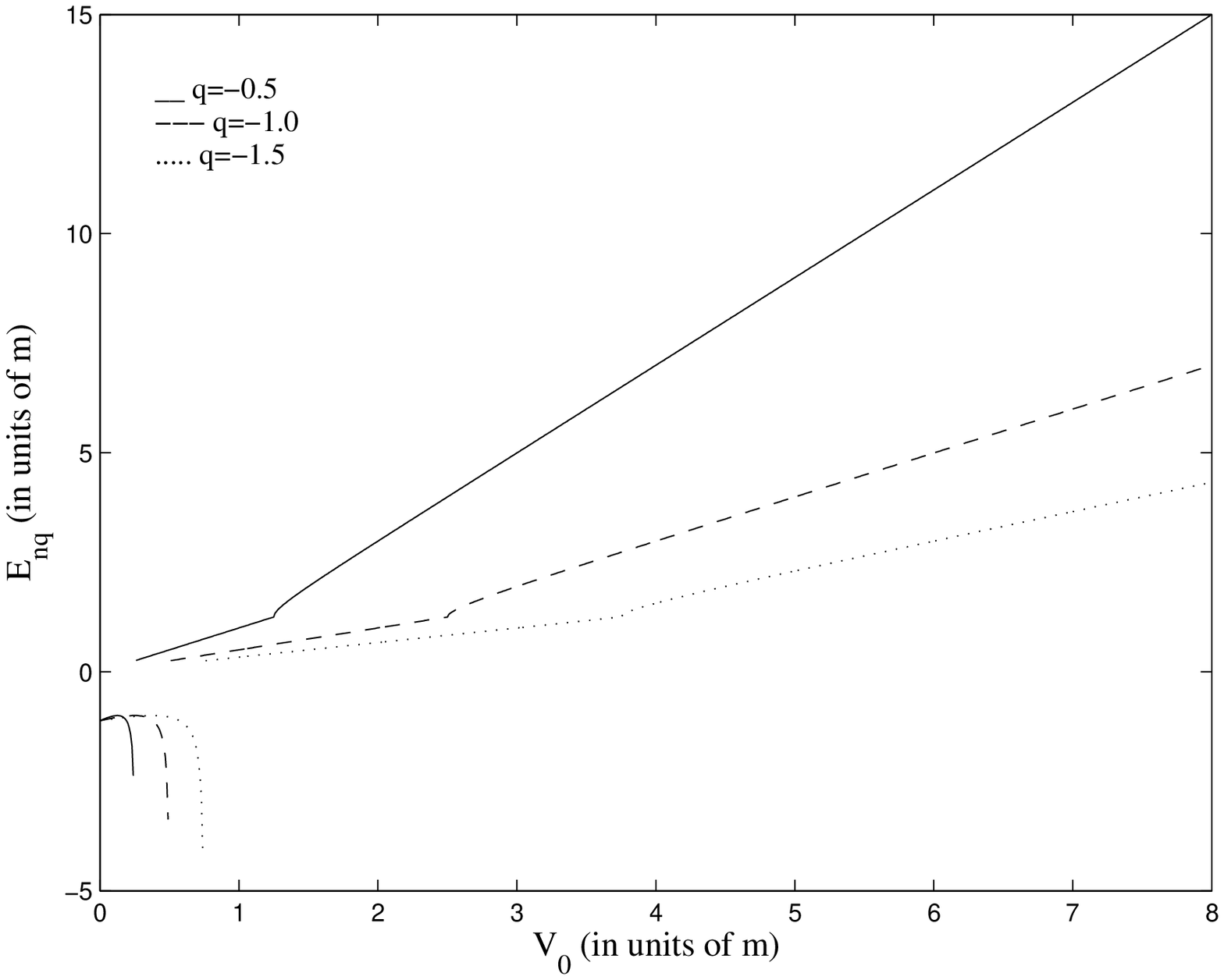}
\end{figure}

\begin{figure}[htbp]
\centering
\includegraphics[height=5in, width=7in]{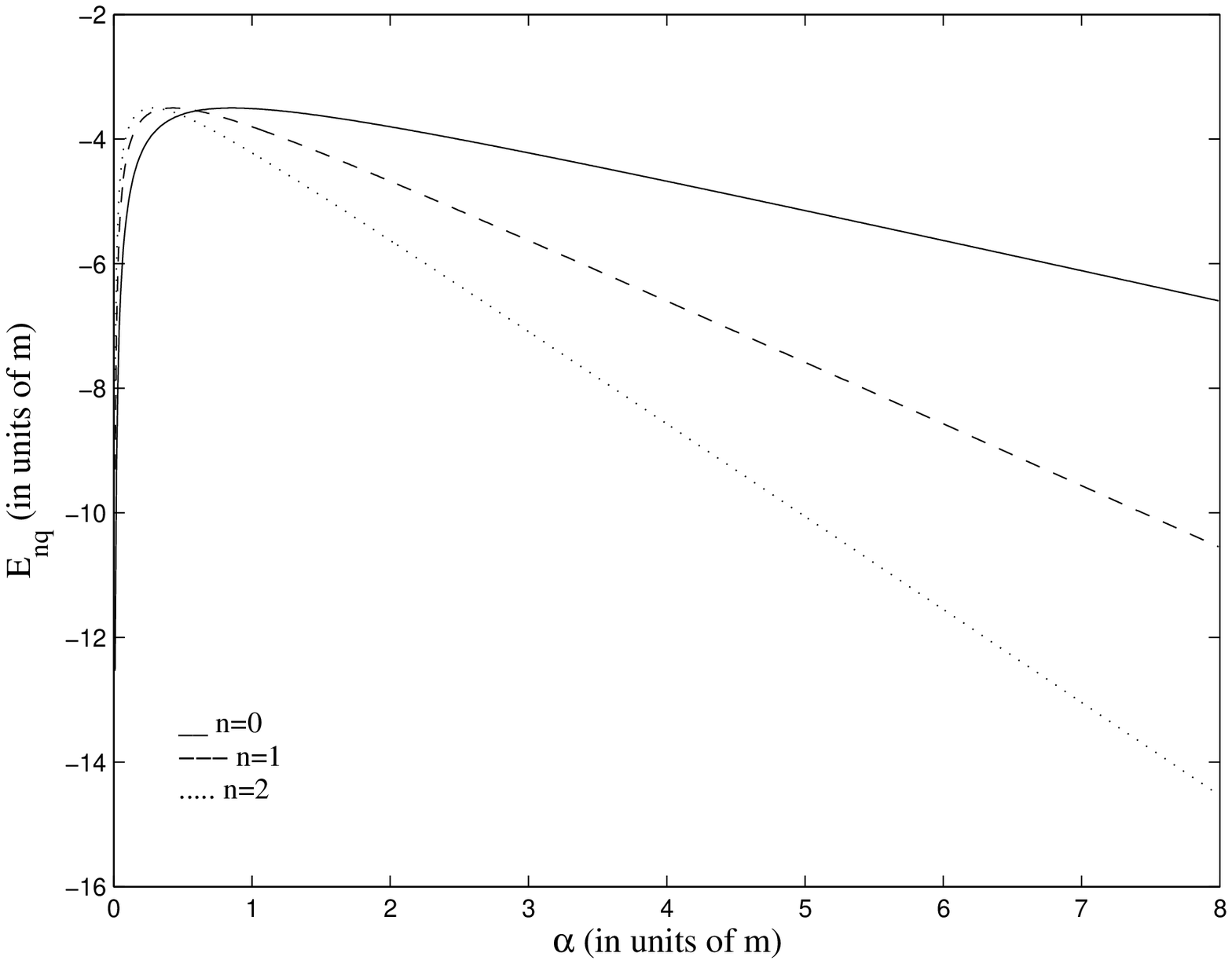}
\end{figure}

\begin{figure}[htbp]
\centering
\includegraphics[height=5in, width=7in]{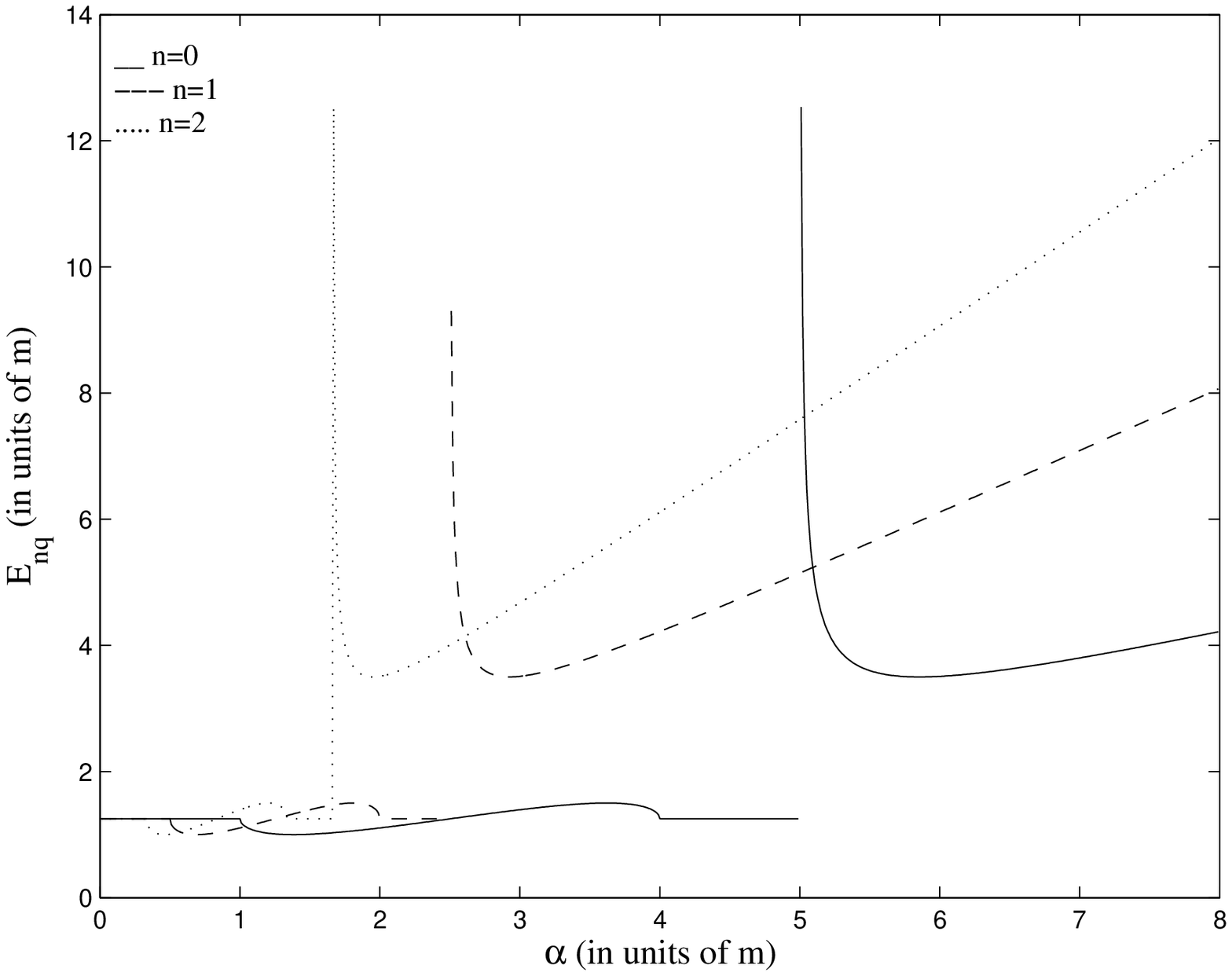}
\end{figure}

\end{document}